\documentclass[12pt]{iopart}
\usepackage{iopams}  
\usepackage{graphicx}
\graphicspath{ {figs/} }

\begin{document}

\title{Tensor Network Finite-Size Scaling for Two-Dimensional 3-state Clock Model}

\author{Debasmita Maiti$^1$, Sing-Hong Chan$^1$, Pochung Chen$^{1,2,3}$}

\address{$^1$ Department of Physics, National Tsing Hua University, Hsinchu 30013, Taiwan}
\address{$^2$ Physics Division, National Center for Theoretical Sciences, Taipei 10617, Taiwan}
\address{$^2$ Frontier Center for Theory and Computation, National Tsing Hua University, Hsinchu 30013, Taiwan}

\ead{pcchen@phys.nthu.edu.tw}

\begin{abstract}
  We benchmark recently proposed tensor network based finite-size scaling analysis 
  in Phys. Rev. B {\bf 107}, 205123 (2023) against two-dimensional classical 3-state clock model.
  Due to the higher complexity of the model, more complicated crossover behavior is observed.
  We advocate that the crossover behavior can be understood from the perspective of finite bond dimension inducing relevant perturbation.
  This leads to a general strategy to best estimate the critical properties for a given set of control parameters.
  For the critical temperature $T_c$, the relative error at the order of $10^{-7}$ can be reached with bond dimension $D=70$.
  On the other hand, with bond dimension $D=60$, the relative errors of the critical exponents $\nu, \beta, \alpha$ are at the order of $10^{-2}$.
  Increasing the bond dimension to $D=90$, these relative errors can be reduced at least to the order of $10^{-3}$.
  In all cases our results indicate that the errors can be systematically reduced  by increasing the bond dimension and the stacking number.
\end{abstract}

%
\vspace{2pc}
\noindent{\it Keywords}: Finite-size scaling, Tensor network renormalization, 3-State clock model

%
\submitto{\NJP}
%
%
%

\section{\label{sec:intro}Introduction}

Tensor network methods have emerged as a powerful tool to study classical and quantum phase transitions \cite{Okunishi.2022}. For two-dimensional (2D) classical models, by representing the partition function as a tensor network and performing tensor renormalization, the nature of the phase transitions can be studied \cite{Ueda.2017, Morita.2019, Ueda.2020ezp, Ueda.2020, Ueda.2021x0n}. An advantage of the tensor network study of the classical models is that extremely large system size can be reached. This effectively reaches the thermodynamic limit, where the spontaneous symmetry breaking can occur. Consequently, the nature of the phase can be studied without using finite-size scaling analysis. 
Moreover,  it is possible to identify fixed-point tensor from which critical properties can be extracted \cite{Evenbly:2015csa, Li.2022ybe, Guo.2023qd, Ueda.2014, Gu.2009, Ueda.2023y3l, Cheng.2023}.
The core of these methods is the scheme to renormalize tensors. 
In recent year, various tensor renormalization schemes have been proposed and studied. This includes Levin-Nave tensor renormalization group (LN-TRG)  \cite{Levin:2007ju}, higher-order tensor renormalization group (HOTRG) \cite{Zhao.2010, Xie:2012iy}, corner transfer matrix renormalization group (CTMRG) \cite{Nishino:1997kn, Orus.2009}, tensor network renormalization (TNR) \cite{Evenbly:2015csa} and loop-optimization of tensor network renormalization (loop-TNR) \cite{Yang:2017hj}. 

Due to the complexity of the 2D tensor contraction, it is necessary to introduce at least one control parameter $D$, called bond dimension, to make the calculation trackable. 
For some schemes, multiple bond dimensions are used.
The finite bond dimension is related to the expressivity and entanglement entropy of the tensor network. As a result, the accuracy of the tensor contraction depends on the bond dimension. For extremely large-size calculations, the only length scale is induced by the bond dimension, and it is possible to perform scaling analysis in terms of the bond dimension. \cite{Pirvu.2012, Ueda.2014,Ueda.2020,Ueda.2023}.
Since the entanglement is finite due to the finite bond dimension, this is referred to as the finite-entanglement scaling regime \cite{Tagliacozzo.2008, Pirvu.2012}.
On the other hand, it is also possible to perform conventional finite-size scaling analysis,
provided that the bond dimension is large enough so that the finite bond dimension effect can be neglected \cite{Hong:2019cm, Huang.2020wh, Huang.2023, Hong.2022, Ueda.2023}. 
This is referred to as the finite-size scaling regime.
In Ref.~\cite{Huang.2023} some of us propose a scheme to perform tensor network based finite-size scaling analysis for the 2D Ising model.
It is shown that critical temperature and critical exponents can be accurately estimated.
More importantly, it is possible to systematically improve the results by increasing the bond dimension.

In this work, we generalize the idea in Ref.~\cite{Huang.2023} to a general scheme 
to determine the critical temperature and critical exponents without a priori knowledge.
Furthermore, we demonstrate that it is important to preserve the symmetry during tensor renormalization,
especially for the estimation of critical exponent associated with the order parameter.
Specifically, we benchmark against the $3$-state clock model.
It is well known that $q$-state clock \cite{Jose.1977} and Potts models \cite{Wu.1982} are two families of important classical statistical models,
which exhibit many of fascinating features of critical phenomena. 
These models serve an important role to study critical phenomena in not only condensed matter physics, 
but also in high energy physics and bio physics~\cite{Potts.1952, Wu.1982}. 
For $q=2$ both models are equivalent to the Ising model,
while for $q=3$ they are the same after a rescaling of the coupling constant \cite{Wu.1982}. 
They have been extensively investigated both analytically and numerically~\cite{Chen:2017ums,Potts.1952,Brito.2007,Brito.2010,Baek.2010,Tomita:2002cka,Tobochnik.1982}. 
Moreover, in recent years  tensor network techniques have been used to study various aspects of the q-state clock and Potts models
~\cite{Morita.2019,Chen:2017ums,Li.20201m7t,Ueda.2020ezp,Ueda.2020,Chen.2011}.
Here we investigate 3-state ferromagnetic clock model by using HOTRG method employing finite-size scaling. 
While a more complex crossover behavior is observed,
it is shown later in this work that by using the general scheme described below, accurate results can be obtained.

In the following we briefly describe our main strategy.
In general we consider some physical quantity $Q$. 
It is expected that $Q$ satisfies a finite-size scaling of the form
\begin{equation}
  \label{eq:FSS}
  Q(L_y) = Q(\infty) + a_0 L_y^{-\omega_0} + a_1 L_y^{-\omega_1} + \cdots,
\end{equation}
where $L_y$ is the system size, $Q(\infty)$ is its thermodynamical value, 
$a_i$s are non-universal constants, and $\omega_0<\omega_1<\cdots$ represent the leading and higher order corrections to the scaling.
Typically one estimates $Q(\infty)$ either by fitting $Q(L_y)$ to Eq.~\ref{eq:FSS} or by using $Q(L_y)$ at the largest $L_y$ available.
If the quantity $Q(L_y)$ is evaluated exactly, one expects that $Q$ approaches $Q(\infty)$ monotonically.
On the other hand, for tensor renormalization based calculations,
typically one finds that initially $Q$ flows toward $Q(\infty)$  monotonically but it start to flow away from $Q(\infty)$ after certain length scale.
From the real-space renormalization group perspective,
we attribute this to the finite bond dimension induced relevant perturbation \cite{Ueda.2023}.
In the beginning the relevant perturbation is weak and the system flows toward the fixed point as one increases the system size.
However, during the renormalization the relevant perturbation also becomes stronger.
Eventually the system starts to flow away from the fixed point.

Our main strategy is to identify the perihelion of the renormalization trajectory, 
i.e., the system size at which the system is closest to the fixed point. We find it natural to consider $\frac{dQ}{d\ln L_y}$.
For exact calculation without relevant perturbation one expects that the magnitude of $\frac{dQ}{d\ln L_y}$ decreases monotonically to zero as system size increases.
In the presence of relevant perturbation, however, $\frac{dQ}{d\ln L_y}$ decreases monotonically to a minimal value at certain length scale $L^*_y$, then it starts to increase.
We argue that $L^*_y$ corresponds to the perihelion of the renormalization trajectory of $Q$ and we use $Q(L^*_y)$ as the best estimation of $Q$.

In practice we first determine the critical  temperature $T_c$ by using the crossing point of the dimensionless $\xi/L_y$,
where $\xi$ is the correlation length.
It is known that the crossing point of dimensionless quantity converges faster to $T_c$ due to the cancellation of the leading order correction.
Our results indicate that $T_c$ can be determined with extremely high accuracy.
On the other hand, there are two types of scaling which can be used to estimate the critical exponents.
First type of method relies on the temperature derivative of some physical quantity, evaluated at a sequence of temperatures.
The requirement is that the temperature sequence shall converge to the critical temperature.
A common sequence to use is the sequence of finite-size crossing point.
The advantage is that a priori knowledge of exact $T_c$ is not required.
Second type of method relies on evaluating some physical quantity at $T_c$.
Since extremely high accuracy can be reached for $T_c$, 
we propose that for both types of methods one can perform the scaling analysis at the above mentioned best estimation of $T_c$.
However, to better understand the intrinsic error and the error due to non-exact $T_c$,
in this work we first estimate the critical exponents by sitting at exact $T_c$.
But we also perform the same scaling analysis with  non-exact $T_c$.
Our results show that the error due to non-exact $T_c$ is much smaller than other intrinsic error.

Finally we would also like to comment on the effect of corner double-line (CDL) tensors before presenting our results.
It is known that CDL tensors contain only ultra-short-range correlator \cite{Gu.2009}.
However, some tensor renormalization methods such as LN-TRG and HOTRG cannot filter out CDL tensors \cite{Gu.2009, Ueda.2014}. 
Conceptually this presents a problem if one is seeking a fixed point tensor. 
The existence of CDL tensors means that a phase is described by a class of equivalent fixed-point tensors instead of by a unique fixed-point tensor.
However, since our approach is based on the finite-size scaling behavior of physical quantities this is not a problem \cite{Kennedy.2022}.
In practice the existence of CDL tensors indicates that part of the computational resources is used to represent and process CDL tensors.
For tensor renormalization methods which can filter out the CDL tensors, it is likely that the same accuracy can be reached with smaller bond dimension. 
However, these methods are also more complex computationally.
Hence, it remains to be investigated which tensor renormalization method can lead to best results within the same limit of computational resource.

The rest of the manuscript is organized as follows: 
In Sec.~\ref{sec:model} we briefly describe the model and summarize the exact results for critical properties.
The information of the construction and renormalization of the tensor is also given.
In Sec.~\ref{sec:result} we present our numerical results for the critical temperature $T_c$ and critical exponents $\nu$, $\beta$, $\alpha$ and crossover length scale. 
We then conclude and provide some further discussion in Sec.~\ref{sec:summary}.

\section{Model and method\label{sec:model}}

We consider 2D classical ferromagnetic 3-state clock model on a square lattice with the Hamiltonian
\begin{equation}
  H_{\mathrm{clock}}[\{\theta\}] = -J\sum_{\langle i,j\rangle} \cos(\theta_{i}-\theta_{j}),
\end{equation}
where $J>0$, $\theta_i \in \{0, \frac{2\pi}{3}, \frac{4\pi}{3}\} $, and $\langle i,j\rangle$ denotes nearest neighbors.
The model exhibits a continuous phase transition from the high-temperature disordered phase to the low-temperature ordered phase.
The exact critical temperature and critical exponents are known,
with $k_B T_c=\frac{3}{2} \frac{1}{\ln(1+\sqrt{3})}\approx1.49246$, 
$\nu=\frac{5}{6}=0.8\bar{3}$, $\beta=\frac{1}{9}=0.\bar{1}$.
and $\alpha=\frac{1}{3}=0.\bar{3}$ \cite{Wu.1982}.
For the convenience of the finite-size scaling analysis, it is also convenient to quote following values:
$\frac{1}{\nu}=\frac{6}{5}=1.2$, $\frac{\beta}{\nu}=\frac{2}{15}=0.1\bar{3}$, $\frac{\alpha}{\nu}=\frac{2}{5}=0.4$ , and $\frac{1-\beta}{\nu}=\frac{16}{15}=1.0\bar{6}$.

In this work, we apply the tensor network based finite-size scaling analysis proposed in Ref.~\cite{Huang.2023}
to determine the critical temperature and the critical exponents.
We start from the tensor network representation of the partition function
\begin{equation}
  Z=\sum_{\{\theta\}} e^{-\beta H[\{\theta\}]}= \mathrm{tTr} \prod_{\mathrm{sites}} \mathbf{T}_{ijkl},
\end{equation}
where $\beta$ is the inverse temperature and tTr is the tensor trace. $\mathbf{T}_{ijkl}$, where $i, j, k, l \in \{0, 1, 2\}$, is the local tensor.
For the 3-state clock model, the local tensor reads \cite{Li.20201m7t}
\begin{equation}
  \mathbf{T}_{ijkl}=  \frac{\sqrt{\lambda_i \lambda_j \lambda_k \lambda_l}}{3} \delta_{\mathrm{mod}(i+j-k-l, 3)},
\end{equation}
where
\begin{equation} 
  \lambda_m=\sum_{\theta_i} \cos(m \theta_i) e^{\beta J \cos ( \theta_i)}.
\end{equation}
Following Ref~\cite{Huang.2023}, we first perform exact contraction of $2\times 2$ or $3\times 3$ sites of $\mathbf{T}$ tensors to construct a $\mathbf{T}^{(d,0)}$ tensor, where $d = 2,3$ is the initial block size. Then we use HOTRG to coarse-grain the tensor iteratively.
After $N$ iterations, one obtains the renormalized tensor $\mathbf{T}^{(d,N)}$ 
which represents approximately $L\times L$ sites of $\mathbf{T}$ tensors where $L=d*2^N$.
By stacking $n$ copies of the $\mathbf{T}^{(d,N)}$ tensors vertically and tracing out the vertical bonds, one obtains the transfer matrix $\mathcal{T}_{L, L_y}$.
It represents a column-to-column transfer matrix with $L_y=nL$ sites, raised to the power of $L$.
In the following we denote the eigenvalue and the eigenvector of $\mathcal{T}_{L, L_y}$ as $\lambda_i (L,L_y)$ and $|\lambda_i (L,L_y)\rangle$ respectively, where $\lambda_i$ is sorted in descending order. By defining $E_i(L_y) = -\ln(\lambda_i (L,L_y))/L$, the correlation length associated with an infinite strip with width $L_y$ can be evaluated as $\xi=\frac{1}{E_1-E_0}$.

The 3-state clock model possesses a $\mathbb{Z}_3$ symmetry.
Due to the symmetry, the magnetization is exactly zero for any finite $L_y$.
At critical temperature, however, it is possible to define a pseudo spontaneous magnetization $m$ \cite{Yang.1952}. 
First define an impurity tensor with following elements 
\begin{equation}
  \mathbf{T}^{z\pm}_{ijkl} = \frac{\sqrt{\lambda_i \lambda_j \lambda_k \lambda_l}}{3} \delta_{\mathrm{mod}(i+j-k-l \pm 1, 3)},
\end{equation}
and renormalize it alongside $\mathbf{T}$ to obtain the renormalized  $\mathbf{T}^{(d,N)}_z$.
Next, one constructs a $\mathcal{T}^{z \pm}_{L,L_y}$ matrix by stacking $n-1$ copies of the $\mathbf{T}^{(d,N)}$ tensors 
and one $\mathbf{T}^{(d,N)}_{z\pm}$ vertically and tracing out the vertical bonds.
The pseudo spontaneous magnetization $m$ is then defined as the smallest eigenvalue of the matrix
\begin{equation}
  \left( \begin{array}{ccc}
    \langle \lambda_0 | \mathcal{T}^z_{L,L_y} | \lambda_0 \rangle & \langle \lambda_0 | \mathcal{T}^z_{L,L_y} | \lambda_1 \rangle & \langle \lambda_0 | \mathcal{T}^z_{L,L_y} | \lambda_2 \rangle \\
    \langle \lambda_1 | \mathcal{T}^z_{L,L_y} | \lambda_0 \rangle & \langle \lambda_1 | \mathcal{T}^z_{L,L_y} | \lambda_1 \rangle & \langle \lambda_1 | \mathcal{T}^z_{L,L_y} | \lambda_2 \rangle \\
    \langle \lambda_2 | \mathcal{T}^z_{L,L_y} | \lambda_0 \rangle & \langle \lambda_2 | \mathcal{T}^z_{L,L_y} | \lambda_1 \rangle & \langle \lambda_2 | \mathcal{T}^z_{L,L_y} | \lambda_2 \rangle
  \end{array} \right),
\end{equation}
where $\mathcal{T}^z \equiv \mathcal{T}^{z+} + \mathcal{T}^{z-}$.

Due to the $\mathbb{Z}_3$ symmetry, the transfer matrix $\mathcal{T}_{L, L_y}$ is block diagonal and can be decomposed into three blocks 
with $\mathbb{Z}_3$ charge $Q=0,1,2$ respectively.
It turns out that $ |\lambda_0 \rangle, |\lambda_1 \rangle, |\lambda_2 \rangle$ have $Q=0,1,2$ respectively and $E_1$ and $E_2$ are degenerate.
Since $\mathcal{T}^z_{L,L_y}$ changes $\mathbb{Z}_3$ charge by $\pm 1$, 
the diagonal elements $ \langle \lambda_i | \mathcal{T}^z_{L,L_y} | \lambda_i \rangle$ should be zero, if symmetric eigenvectors are used.
However, if symmetry is not enforced during the tensor renormalization, 
numerical inaccuracy and truncation may lead to $\mathbb{Z}_3$ symmetry breaking and diagonal elements become non-zero.
Based on these observations we use the sum $ \sum_{q=0,1,2}  \langle \lambda_q | \mathcal{T}^z_{L,L_y} | \lambda_q \rangle $ 
as a measure to gauge the effect of these unintentional symmetry breaking.
As it is shown later that this quantity may suddenly become larger after some tensor renormalization 
and make the calculation of the magnetization unstable.

\section{Numerical Results and Finite-Size Scaling Analysis\label{sec:result}}

\subsection{Critical temperature $T_c$}

\begin{figure}[t]
  \centering
  \includegraphics[width=0.95\columnwidth]{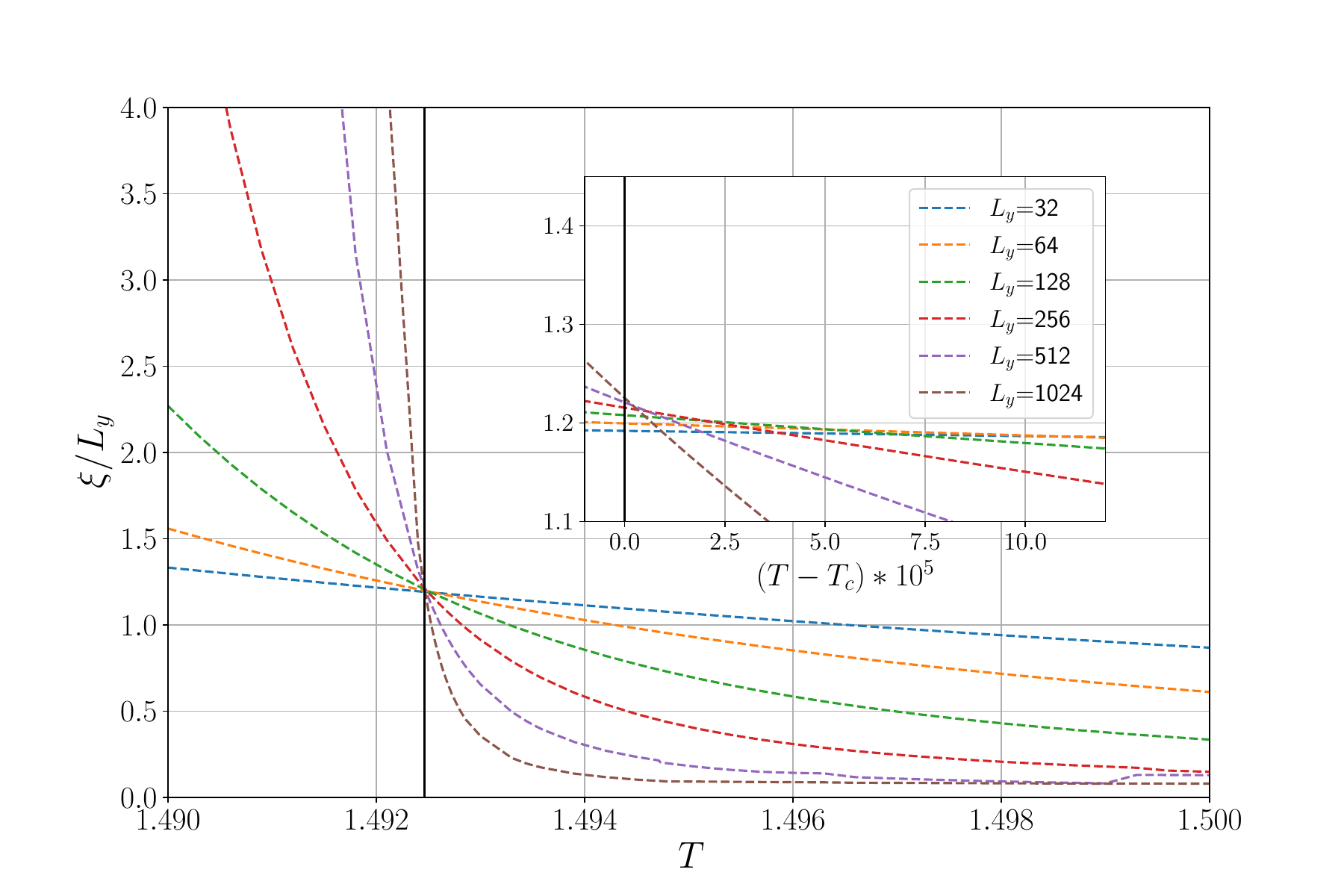}
  \caption{\label{fig:xi_over_L}
   $\xi/L_{y}$ as a function of $T$ with various $L_y$, obtained with $D=70$ and $n=3$.
   Inset: Zoomed plot near $T_c$.}
\end{figure}

\begin{figure}[t]
  \centering
  \includegraphics[width=0.95\columnwidth]{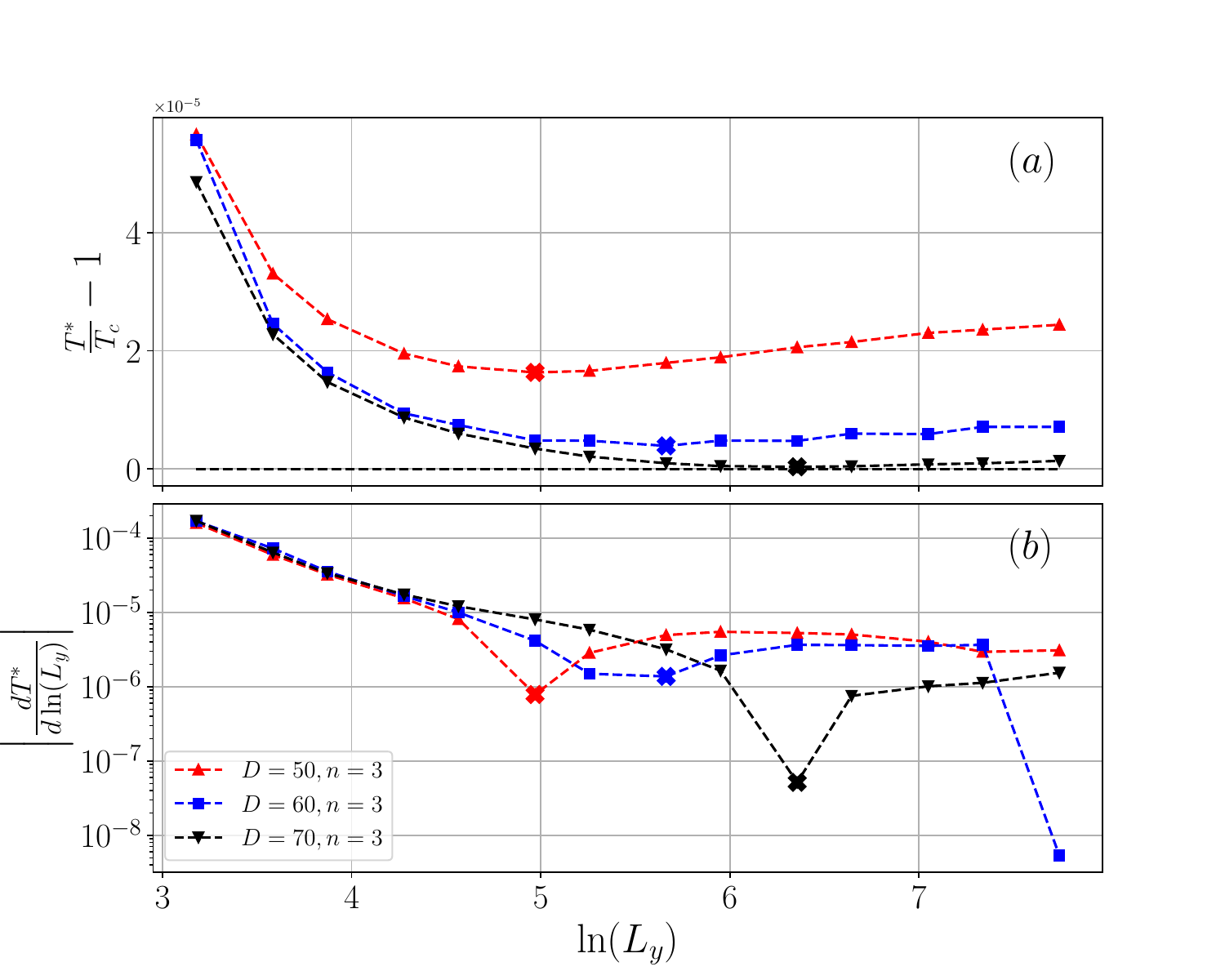}
  \caption{\label{fig:Ts_xi_over_L}  
  (a) $\frac{T^*}{T_c}-1$ as a function of $\ln(L_y)$. 
  Horizontal dashed line corresponds to the exact value. Dashed lines connecting symbols are guides for the eye.
  (b) $\left| \frac{dT^*}{d\ln(L_y)} \right|$ as a function of $\ln(L_y)$.
  The crosses indicate the $L^*_y$ where $\left| \frac{dT^*}{d\ln(L_y)} \right|$ attains its first local minimum.}
\end{figure}

Following conventional finite-size scaling analysis, we first use the crossing point of the dimensionless $\xi/L_y$ to estimate the critical temperature $T_c$. 
In Figure~\ref{fig:xi_over_L}, $\xi/L_y$ is plotted as a function of the temperature $T$ for various $L_y$, where $D=70$ and $n=3$ are used.
As expected, curves from different sizes cross near $T_c$ which is indicated by the vertical black line.
In the inset, we zoom in near $T_c$ and the drifting of the crossing point is clearly observed.
It is common to define the crossing point of $\xi(L_y)/L_y$ and $\xi(2L_y)/(2L_y)$ as $T^*(L_y)$, 
which represents the finite-size estimation of $T_c$ and it approaches $T_c$ as $L_y \rightarrow \infty$.
Based on our general strategy, we plot the relative error $\frac{T^*}{T_c}-1$ and $\left| \frac{dT^*}{d\ln(L_y)} \right|$ 
respectively as a function of $\ln(L_y)$ in Figure~\ref{fig:Ts_xi_over_L}(a) and (b).
Here only results from $D=50,60,70,$ and $n=3$ are plotted for clarity.
It should be noted that the labels for y-axis in Figure~\ref{fig:Ts_xi_over_L}(a) is at the order of $10^{-5}$.
We observe that $T^*$ flows toward $T_c$ for smaller $L_y$, but it starts to flow away for larger $L_y$.
Similarly, we find that $\left| \frac{dT^*}{d\ln(L_y)} \right|$ decreases monotonically toward zero for smaller $L_y$, 
but for larger $L_y$ it shows non-monotonic behavior.
This can be attributed to the crossover between finite-size scaling regime and finite-entanglement scaling regime \cite{Huang.2023}.
As discussed in the introduction, we define $L^*_y$ to be the size at which $\left| \frac{dT^*}{d\ln(L_y)} \right|$ attains its first local minimum as $L_y$ increases
and use $T^*(L^*_y)$ as the best estimation of $T_c$ for a fixed $D$ and $n$.
We emphasize that the procedure does not require a priori knowledge of exact $T_c$.
The dashed line in  Figure~\ref{fig:Ts_xi_over_L}(a),
which corresponds to the exact value, is plotted only as a reference.
Moreover, we find that $T^*(L_y)$ is quite flat near $L^*_y$ and it would be desirable to have more data near $L^*_y$ to obtain a better estimation.
However, one cannot evaluate $T^*$ at all possible $L_y$ due to the nature of HOTRG.
Consequently the $L^*_y$ obtained here only represents a rough estimation of the crossover length scale.

In Table~\ref{tab:Ts_xi_over_L},  we summarize our results of the $L^*_y$ and the relative error $ \frac{T^*}{T_c}-1 $. 
We find that the error can be systematically improved by increasing $D$ or $n$.
With $D=50$ and $n=1$, the relative error is already at the order of $10^{-5}$ which is quite small.
With $D=70$ and $n=3$, the relative error can be further reduced to the order of $10^{-7}$.
Our results indicate that $T_c$ can be estimated with extremely high accuracy.
Finally we note that results from $\mathbb{Z}_3$ symmetric HOTRG are presented here.
We have also performed the calculations in which the $\mathbb{Z}_3$ symmetry is not explicitly preserved.
We find that the identification of $L^*_y$ remains the same and relative error is of the same order.

\begin{table}
\caption{\label{tab:Ts_xi_over_L}Summary of the relative error $ \epsilon_T \equiv  \frac{T*}{T_c}-1 $.}
\begin{indented}
\item[]\begin{tabular}{@{}lllllll}
\br
$D$  & $L_y^*$&$ \epsilon_T $&$L_y^*$&$ \epsilon_T $&$L_y^*$&$\epsilon_T $\\ 
\mr
$50$ & 256     & $+1.97\times 10^{-5}$  & 192 & $+1.83\times 10^{-5}$ & 144 & $+1.63\times 10^{-5}$\\  
$60$ & 512  & $+6.73\times 10^{-6}$  & 256 & $+4.72\times 10^{-6}$ & 288 & $+3.91\times 10^{-6}$\\     
$70$ & 768     & $+6.51\times 10^{-7}$  & 512 & $-2.64\times 10^{-9}$ & 576 & $+3.70\times 10^{-7}$
\end{tabular}
\end{indented}
\end{table}

\subsection{Critical exponent $\nu$}

\begin{figure}[t]
  \centering
  \includegraphics[width=0.95\columnwidth]{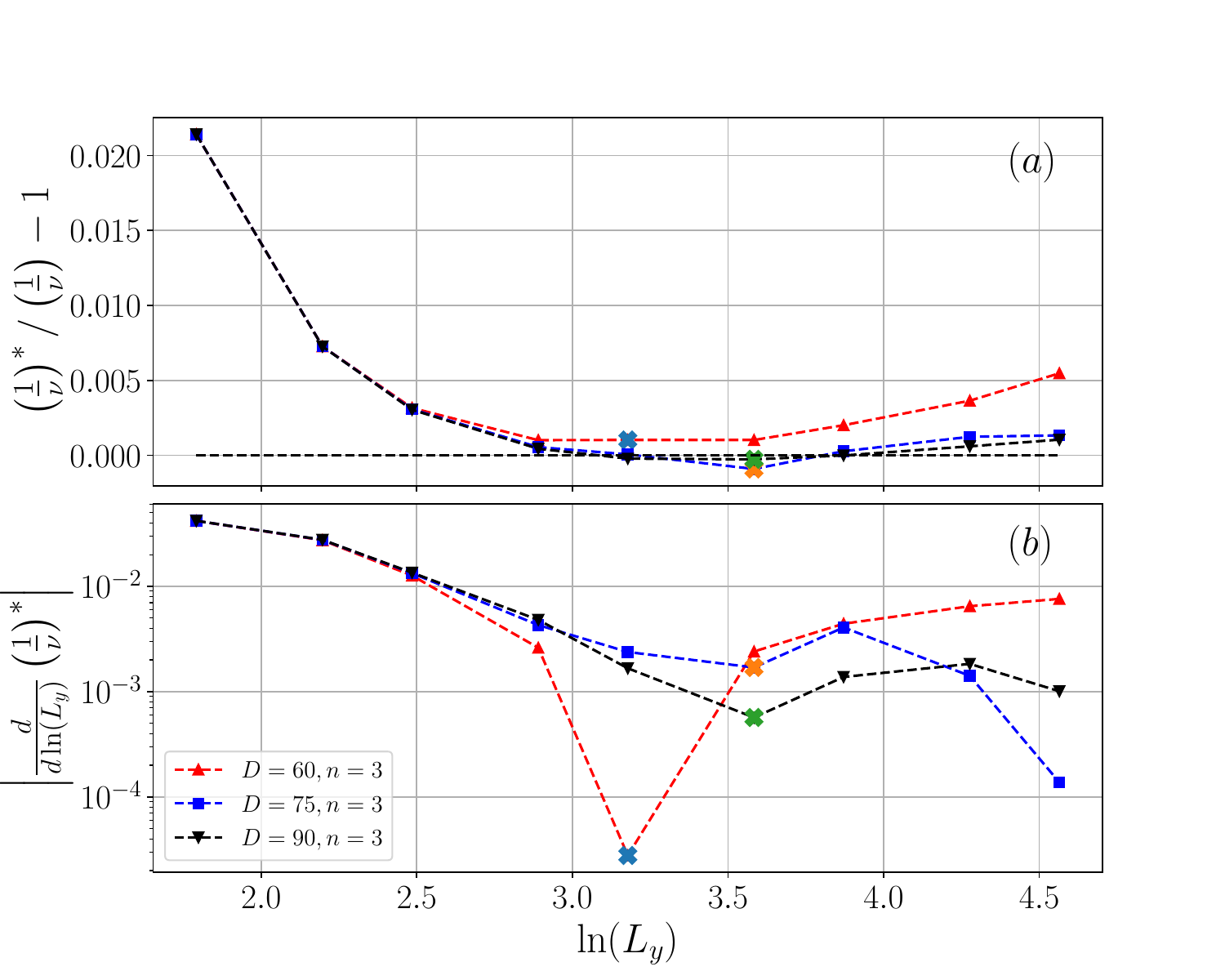}
   \caption{\label{fig:nu} 
   (a) $\left(\frac{1}{\nu}\right)^*/\left(\frac{1}{\nu}\right)-1$ as a function of $\ln(L_y)$. 
    Horizontal dashed line corresponds to the exact value. Dashed lines connecting symbols are guides for the eye.
   (b) $\left| \frac{d}{d\ln(L_y)}\left(\frac{1}{\nu}\right)^* \right|$ as a function of $\ln(L_y)$.
  The crosses indicate the $L^*_y$ where $\left| \frac{d}{d\ln(L_y)}\left(\frac{1}{\nu}\right)^* \right|$ attains its first local minimum.}
\end{figure}

To estimate the critical exponent $\nu$, we first define the finite-size estimation of $\frac{1}{\nu}$ as
\begin{equation}
  \label{eq:nu}
  \left( \frac{1}{\nu} \right)^* \equiv \frac{\ln S_{\xi/L_y}(2L_y)-\ln S_{\xi/L_y}(L_y)} {\ln(2)},
\end{equation}
where $S_{\xi/L_y} \equiv \left. \frac{d}{dT}(  \frac{\xi}{L_y} ) \right|_{T_c}$ is the slope of $\frac{\xi}{L_y}$ at $T_c$ {\cite{Huang.2023}}.
Based on the finite-size scaling theory, it is expected that $\left( \frac{1}{\nu} \right)^*$ should approach the exact value $\frac{1}{\nu}=\frac{6}{5}$ in the thermodynamic limit. 
Following again the general strategy,
in Figure~\ref{fig:nu}(a) we plot the relative error $\left( \frac{1}{\nu} \right)^*/\left( \frac{1}{\nu} \right)-1$ as a function of $\ln(L_y)$,
while in Figure~\ref{fig:nu}(b) we plot $\left| \frac{d (1/\nu)^*}{d\ln(L_y)} \right|$. 
The results follow the general trend as described in the introduction and are similar to the results of $T^*$.
To identify the perihelion of the renormalization trajectory of $\left( \frac{1}{\nu} \right)^*$, we locate the size $L^*_y$ at which $\left| \frac{d (1/\nu)^*}{d\ln(L_y)} \right|$ attains its first local minimum as $L_y$ increases.
Then $\left( \frac{1}{\nu} \right)^*$ at $L^*_y$ is used as the best estimation of $\frac{1}{\nu}$ for a fixed set of parameters.
In Table~\ref{tab:nu} we summarize our results for the relative error.
With $D=60$ and $n=1$, the relative error is about $10^{-3}$ which is small.
By using $D=90$ and $n=3$, it can be further reduced to the order of $10^{-4}$.
An interesting observation is that the $L^*_y$ identified here is very different from the one identified via $T^*$.
As one sees from the results of other critical exponents, the value and the trend of $L^*_y$ are different for different physical quantity.
Finally we note that the results from $\mathbb{Z}_3$ symmetric HOTRG are presented here
and the results from non-symmetry-preserving HOTRG are similar.

Some additional comments are now in order. 
In the estimation of $1/\nu$ above, we evaluate the slope in Eq.~\ref{eq:nu} at exact $T_c$.
This is because we would like to investigate the intrinsic error due to the tensor renormalization.
In general the exact $T_c$ is not known.
In this case, we advocate that the best strategy is to estimate the critical exponents based on the quantities evaluated at the best estimated $T_c$, i.e., $T^*(L^*_y)$.
This is based on the observation that $T_c$ can be estimated with extremely high accuracy and the error of the critical exponent is much larger than the error of critical temperature.
To further support this, we have performed the same analysis but with data evaluated at temperature shifted away from $T_c$ by an order of $10^{-5}$ to account for the error in the estimated value of $T_{c}$. 
First, we find that the same $L^*_y$ is identified.
Furthermore, while in general the relative error increases, it typically remains within the same order of magnitude.
These results indicate that the error due to using non-exact $T_c$ is much smaller than the intrinsic error, in support of our claim.

\begin{table}
\caption{\label{tab:nu}Summary of the relative error $\epsilon_\nu \equiv \left(\frac{1}{\nu}\right)^*/\left(\frac{1}{\nu}\right)-1$.} 
\begin{indented}
\item[]\begin{tabular}{@{}lllllll}
\br
$D$  & $L_y^*$ &  $\epsilon_\nu$  & $L_y^*$  &  $\epsilon_\nu$ & $L_y^*$  &  $\epsilon_\nu$\\ 
\mr
$60$ & 16  & $+7.86\times 10^{-3}$ & 24 & $+1.99\times 10^{-3}$  & 24 & $+1.02\times 10^{-3}$\\   
$75$ & 16  & $+3.96\times 10^{-3}$ & 24 & $-4.44\times 10^{-4}$  & 36 & $-9.12\times 10^{-4}$\\     
$90$ & 16  & $+4.69\times 10^{-3}$ & 24 & $+3.07\times 10^{-4}$  & 36 & $-2.70\times 10^{-4}$  
\end{tabular}
\end{indented}
\end{table}

\subsection{Critical exponent $\beta$}

\begin{figure}[t]
  \includegraphics[width=0.95\columnwidth]{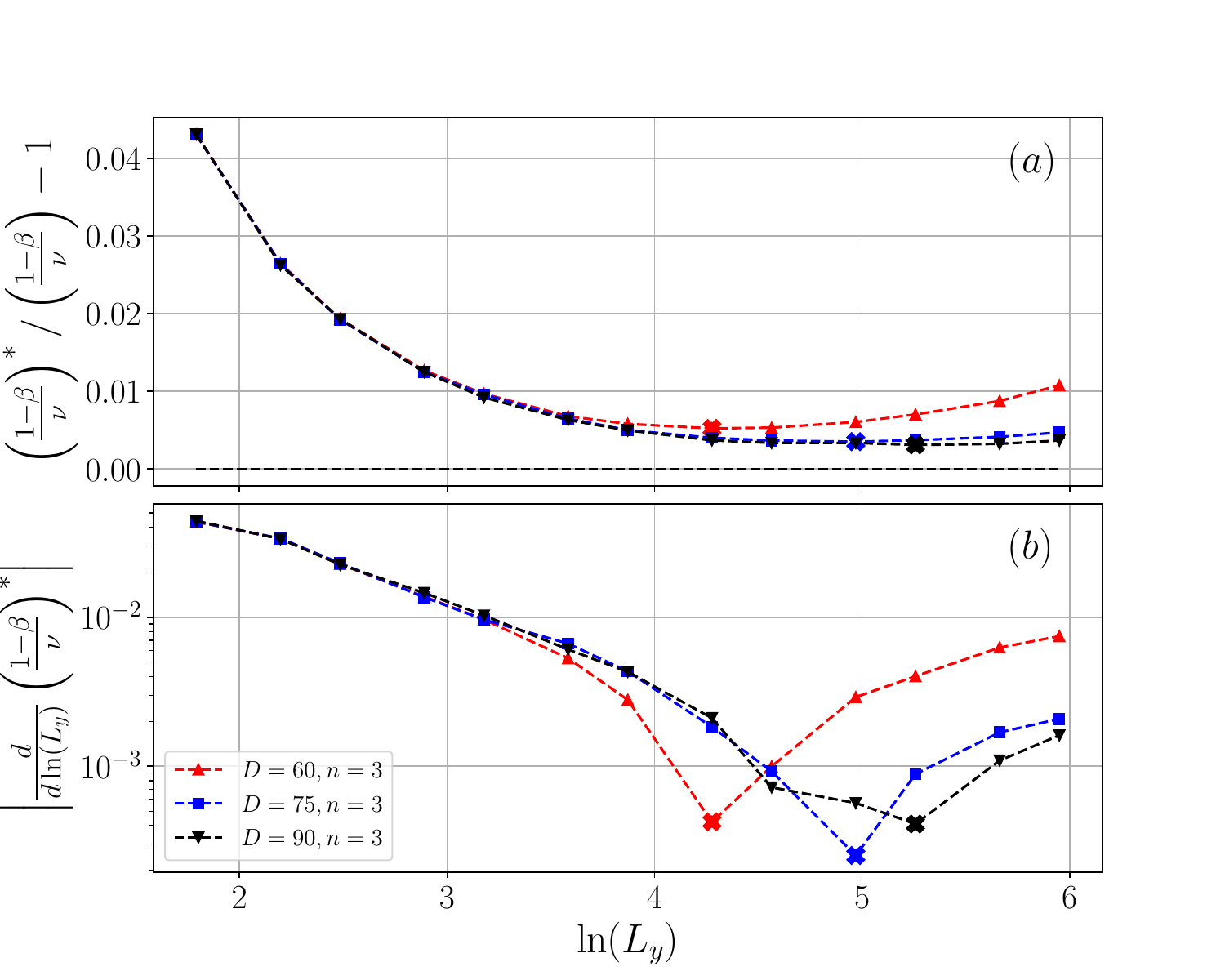}
   \caption{\label{fig:1_beta} 
   (a) $\left( \frac{1-\beta}{\nu} \right)^*/\left( \frac{1-\beta}{\nu} \right) -1 $ as a function of $\ln(L_y)$. 
    Horizontal dashed line corresponds to the exact value. Dashed lines connecting symbols are guides for the eye.   
   (b) $\left| \frac{d}{d\ln(L_y)}\left(\frac{1-\beta}{\nu}\right)^* \right|$ as a function of $\ln(L_y)$.
  The crosses indicate the $L^*_y$ where $\left| \frac{d}{d\ln(L_y)}\left(\frac{1-\beta}{\nu}\right)^* \right|$ attains its first local minimum.   
   }.
 \end{figure}

\begin{figure}[t]
  \includegraphics[width=0.95\columnwidth]{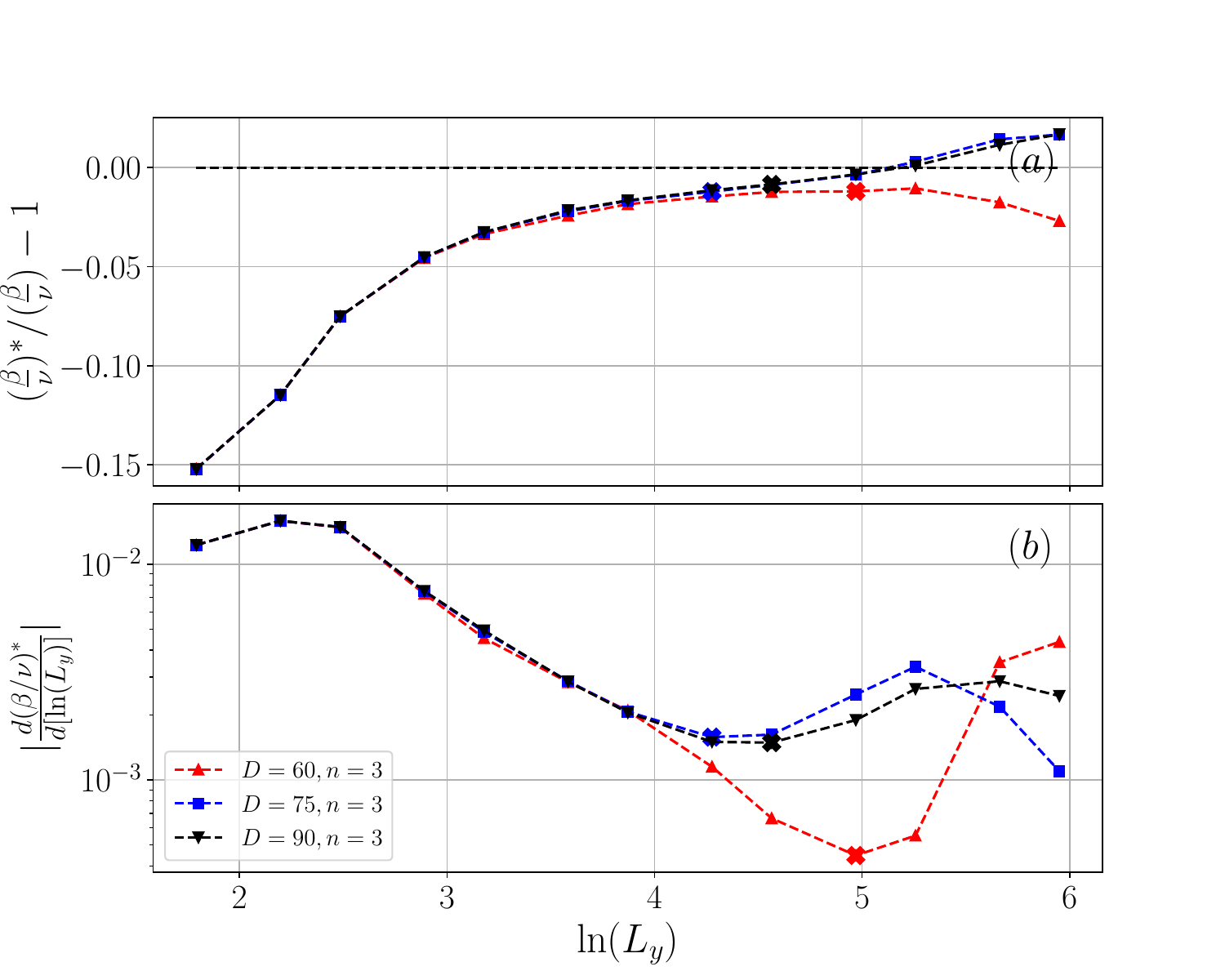}
   \caption{\label{fig:m_Tc} 
   (a) $\left( \frac{\beta}{\nu} \right)^*/\left( \frac{\beta}{\nu} \right)-1$ as a function of $\ln(L_y)$. 
  Horizontal dashed line corresponds to the exact value. Dashed lines connecting symbols are guides for the eye.
   (b) $\left| \frac{d}{d\ln(L_y)}\left(\frac{\beta}{\nu}\right)^* \right|$ as a function of $\ln(L_y)$.
  The crosses indicate the $L^*_y$ where $\left| \frac{d}{d\ln(L_y)}\left(\frac{\beta}{\nu}\right)^* \right|$ attains its first local minimum. 
  }
\end{figure}

In this work we use two  approaches to estimate the critical exponent $\beta$.
First approach uses the slope of the pseudo spontaneous magnetization $m$ at $T_c$ to estimate $ \frac{1-\beta}{\nu} $.
Similar to the finite-size estimation of $\frac{1}{\nu}$, we define the finite-size estimation of $\frac{1-\beta}{\nu}$ as
\begin{equation}
  \left( \frac{1-\beta}{\nu} \right)^* \equiv \frac{\ln S_{m}(2L_y)-\ln S_{m}(L_y)} {\ln(2)},
\end{equation}
where $S_{m} \equiv \left. \frac{dm}{dT} \right|_{T_c}$ is the slope of the $m$ at $T_c$.
Second approach uses the finite-size scaling behavior of $m$ at $T_c$ to estimate $\beta$.
Since one expects $m(T_c, L_y) \propto L_y^{-\beta/\nu}$,
we define the finite-size estimation of $\frac{\beta}{\nu}$ as
\begin{equation}
  \left( \frac{\beta}{\nu} \right)^* \equiv -\frac{d \ln(m(T_c, L_y))}{d \ln(L_y)},
\end{equation}
and  evaluate it numerically. 
In the thermodynamic limit, it is expected that $\left( \frac{1-\beta}{\nu} \right)^* \rightarrow \frac{1-\beta}{\nu}=\frac{16}{15}$ 
and $\left( \frac{\beta}{\nu} \right)^* \rightarrow \frac{\beta}{\nu}=\frac{2}{15}$.
In Figure~\ref{fig:1_beta}(a) and ~\ref{fig:m_Tc}(a) we plot respectively the relative error
$\left( \frac{1-\beta}{\nu} \right)^*/\left( \frac{1-\beta}{\nu} \right)-1 $ and 
$\left( \frac{\beta}{\nu} \right)^*/\left( \frac{\beta}{\nu} \right)-1 $ 
as a function of $\ln(L_y)$,
where the dashed line corresponds to the exact value. 
To identify $L_y^*$,  we plot $\left| \frac{d}{d\ln(L_y)}\left( \frac{1-\beta}{\nu} \right)^* \right|$ in Figure~\ref{fig:1_beta}(b) 
while  $\left|\frac{d \left( \beta/\nu \right)^*}{d \ln(L_y)}\right|$ is plotted in Figure~\ref{fig:m_Tc}(b).
Again we find it follows the general trend as described in the introduction
and we assign $L_y^*$ to be the size at which it attains first local minimum.
We summarize the results in terms of relative error in Table~\ref{tab:beta} and Table~\ref{tab:beta_Tc}. 

\begin{table}
\caption{\label{tab:beta}Summary of the relative error $\epsilon_{(1-\beta)} \equiv \left( \frac{1-\beta}{\nu} \right)^*/\left( \frac{1-\beta}{\nu} \right)-1 $.}
\begin{indented}
\item[]\begin{tabular}{@{}lllllll}
\br
$D$  &  $L_y^*$  &  $\epsilon_{(1-\beta)}$  & $L_y^*$  & $\epsilon_{(1-\beta)}$  & $L_y^*$  & $\epsilon_{(1-\beta)}$   \\
\mr
$ 60 $ & 32   &  $+1.52\times 10^{-2}$ & 64 & $+7.14 \times 10^{-3}$ & 72   & $+5.22 \times 10^{-3}$ \\  
 $ 75 $ & 48   &  $+1.09\times 10^{-2}$ & 96 & $+5.12 \times 10^{-3}$ & 144   & $+3.51 \times 10^{-3}$ \\  
 $ 90 $ & 48   &  $+7.77\times 10^{-3}$ & 128 & $+4.56 \times 10^{-3}$ & 192   & $+3.10 \times 10^{-3}$ \\ 
\end{tabular}
\end{indented}
\end{table}


\begin{table}
\caption{\label{tab:beta_Tc}Summary of the relative error $ \epsilon_{\beta} \equiv \ \left( \frac{\beta}{\nu} \right)^*/\left( \frac{\beta}{\nu} \right)-1 $.}
\begin{indented}
\item[]\begin{tabular}{@{}lllllll}
\br
$D$    & $L_y^*$ &  $\epsilon_{\beta}$   & $L_y^*$  &  $\epsilon_{\beta}$  & $L_y^*$  &  $\epsilon_{\beta}$    \\
\mr
$ 60 $ & 64  &  $-3.28 \times 10^{-2}$   & 192& $-1.16 \times 10^{-2}$   & 144 & $-1.20\times 10^{-2}$  \\   
  $ 75 $ & 64  &  $-2.74 \times 10^{-2}$   &  64 & $-1.50 \times 10^{-2}$   & 72   & $-1.22\times 10^{-2}$  \\    
  $ 90 $ & 64  &  $-2.41 \times 10^{-2}$   &  96 & $-9.97 \times 10^{-3}$   & 96   & $-8.51\times 10^{-3}$   
\end{tabular}
\end{indented}
\end{table}

\begin{figure}
  \includegraphics[width=0.95\columnwidth]{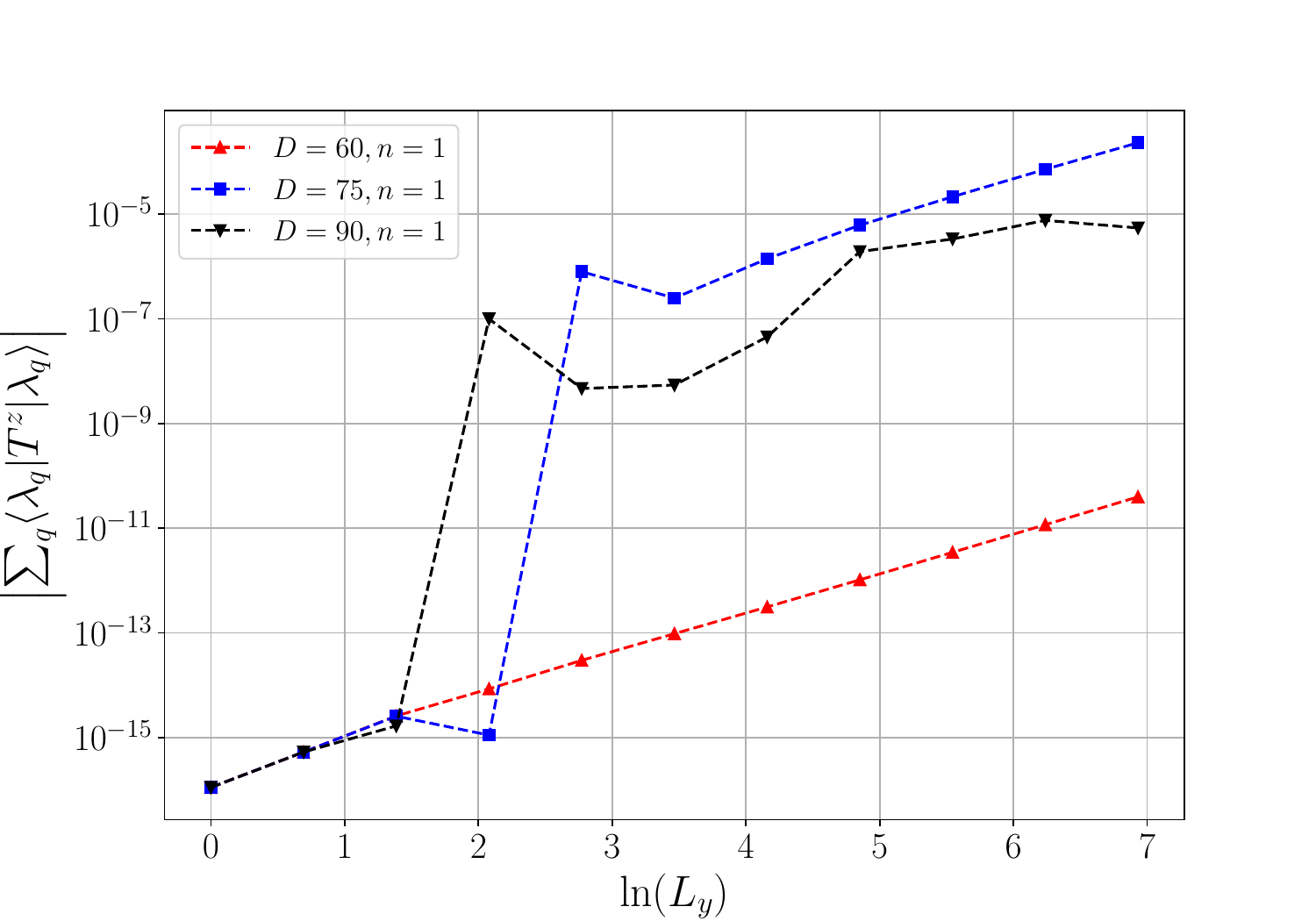}
  \caption{\label{fig:traceM} $\sum_{q}  \langle \lambda_q | \mathcal{T}^z_{L,L_y} | \lambda_q \rangle$ as a function of $\ln(L_y)$.
  Dashed lines connecting symbols are guides for the eye.}
 \end{figure}

The results show that first approach has slightly better accuracy.
The error of second approach is about two to three times larger.
With $D=60$, $n=1$, the relative error is at the order of $10^{-2}$.
Pushing to $D=90$, $n=3$, the relative error is reduced to $10^{-3}$,
and the overall improvement is similar for both methods.
We note that for both approaches results from HOTRG which explicitly preserves the $\mathbb{Z}_3$ symmetry are plotted here.
On the other hand, we find that if the $\mathbb{Z}_3$ symmetry is not explicitly preserved in the HOTRG, 
the estimation of $\left( \frac{1-\beta}{\nu} \right)^*$ becomes unstable for larger $D$
while the estimation of $\left( \beta/\nu \right)^*$ remains stable.
We attribute this instability to the accidental symmetry breaking. 
As mentioned in Sec.~\ref{sec:model}, one can use $\sum_{q}  \langle \lambda_q | \mathcal{T}^z_{L,L_y} | \lambda_q \rangle$ to gauge the accidental symmetry breaking.
In Figure~\ref{fig:traceM} we plot this quantity at $T_c$ as a function of $\ln(L_y)$ on a semilog plot using HOTRG which does not explicitly preserves the $\mathbb{Z}_3$ symmetry.
We observe a smooth linear dependence for $D=60$, but for $D=75, 90$ there is a sudden jump of several order of magnitude at some length scale.
The reason why this happens at larger $D$ and $L_y$ can be understood intuitively.
As $D$ and $L_y$ increase the singular values in the tail also become smaller.
This makes it easier to accidentally mix sectors with different quantum numbers during HOTRG, resulting the accidental symmetry breaking.
Once accidental symmetry breaking occurs, the quantity shown in Figure~\ref{fig:traceM} exhibits a sudden jump.
However, exactly when such an accidental symmetry breaking occurs depends sensitively on various parameters.
Consequently, the numerical derivative $\left. \frac{dm}{dT} \right|_{T_c}$ and hence the estimation of $\left( \frac{1-\beta}{\nu} \right)^*$ becomes numerically unstable at  larger $D$ and $L_y$.
This observation shows that it is essential to explicitly keep the symmetry when estimating the exponent related to the order parameter.

\subsection{Critical exponent $\alpha$}

\begin{figure}
   \includegraphics[width=0.95\columnwidth]{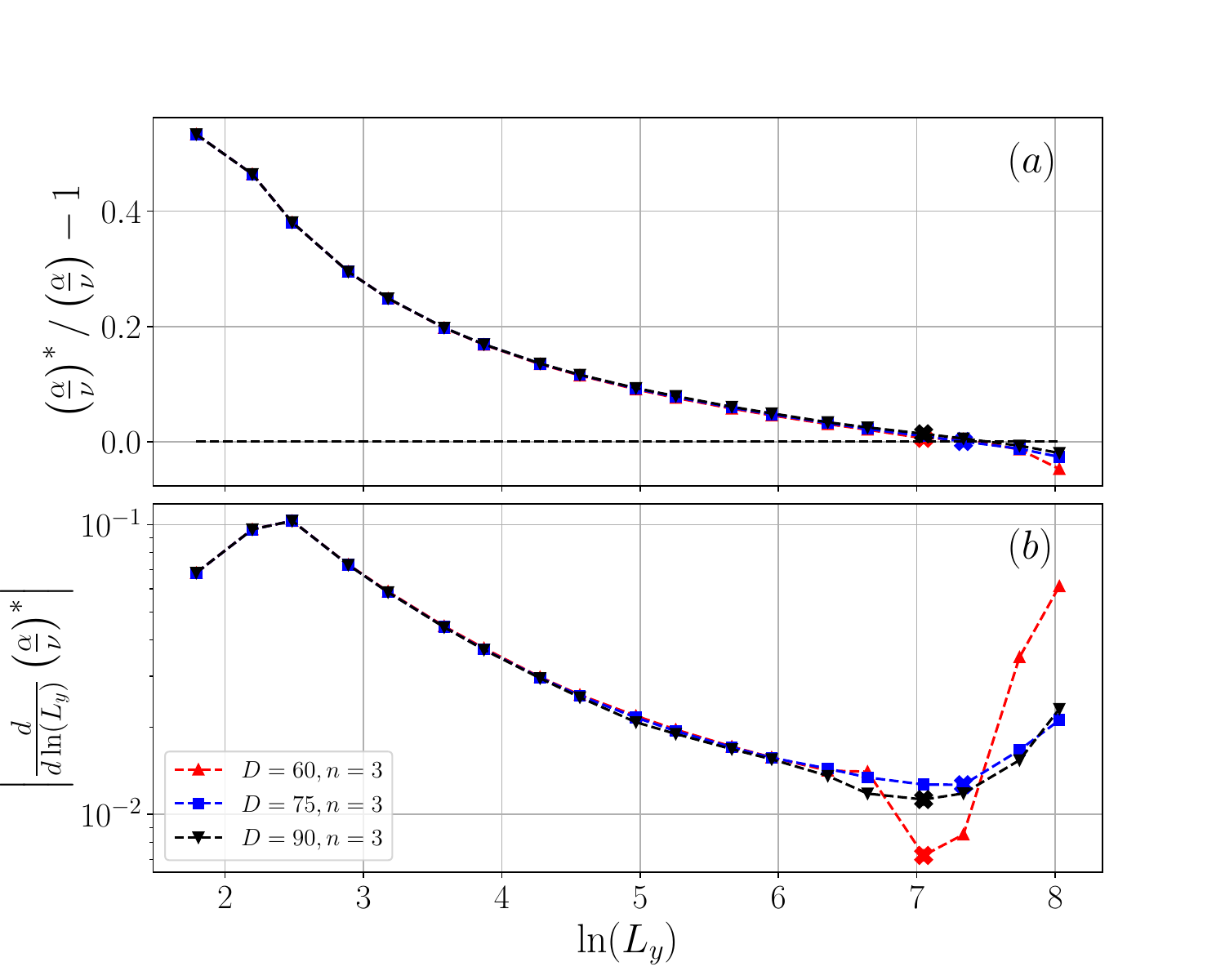}
   \caption{\label{fig:c_v} 
   (a) $\left( \frac{\alpha}{\nu} \right)^*/\left( \frac{\alpha}{\nu} \right)-1$ as a function of $\ln(L_y)$. 
   Horizontal dashed line corresponds to the exact value. Dashed lines connecting symbols are guides for the eye.   
   (b) $\left|\frac{d}{d \ln(L_y)}  \left( \frac{\alpha}{\nu} \right)^* \right|$ as a function of $\ln(L_y)$.
   The crosses indicate the $L^*_y$ where it attains its first local minimum.}
 \end{figure}

To estimate critical exponent $\alpha$,  we start with $\mathbf{T}_{\langle i,j\rangle} \equiv \mathbf{T}^{z+}_i \mathbf{T}^{z-}_j  + \mathbf{T}^{z-}_i \mathbf{T}^{z+}_j$ 
where $i, j$ are nearest neighbours and renormalize it alongside $\mathbf{T}$.      
It is then straightforward to evaluate the energy per bond from the renormalized $\mathbf{T}^{(d,N)}$ and $\mathbf{T}^{(d,N)}_{\langle i,j\rangle}$.
Next we perform numerical derivative to obtain the specific heat per site $c_v$, and the finite-size estimation of $\left(\frac{\alpha}{\nu}\right)$
which is defined as 
\begin{equation}
  \left( \frac{\alpha}{\nu} \right)^* \equiv \frac{d \ln(c_v(T_c, L_y))}{d \ln(L_y)}.
\end{equation}
In Figure~\ref{fig:c_v}(a) we plot the relative error $\left( \frac{\alpha}{\nu} \right)^*/\left( \frac{\alpha}{\nu} \right)-1$ as a function of $\ln(L_y)$, 
where the black dashed line corresponds to the exact value.
We observe that $\left( \frac{\alpha}{\nu} \right)^*$ shows a much stronger finite-size and larger system size is needed to reach similar relative error.
Next we plot $\left|\frac{d}{d \ln(L_y)}  \left( \frac{\alpha}{\nu} \right)^* \right|$ in Figure~\ref{fig:c_v}(b) 
and we use the same strategy to identify the length scale $L_y^*$ and to estimate the exponent $\alpha$.
In Table.~\ref{tab:alpha} we summarize our findings. 

We note that results from $\mathbb{Z}_3$ symmetric calculations are presented here
and calculations without enforcing $\mathbb{Z}_3$ symmetry becomes unstable to determine the exponent for larger $L$ and $D$.
This is because it is necessary to take two numerical steps to obtain $\left( \frac{\alpha}{\nu} \right)^*$
and another one to obtain $\left|\frac{d}{d \ln(L_y)}  \left( \frac{\alpha}{\nu} \right)^* \right|$, they become numerically less stable at larger $L_y$.
Furthermore, while qualitatively the accuracy increases when $D$ or $n$ are increased, it becomes tricky to identify $L_y^*$.
Consequently, best estimated exponents are not monotonic as one increases $D$ or $n$.
However, the accuracy is at least at the order of $10^{-2}$ and can be reduced to $10^{-4}$.

\begin{table}
\caption{\label{tab:alpha}Summary of relative error $\epsilon_{\alpha} \equiv \left( \frac{\alpha}{\nu} \right)^*/\left( \frac{\alpha}{\nu} \right)-1$.}
\begin{indented}
\item[]\begin{tabular}{@{}lllllll}
\br
$D$    & $L_y^*$ &  $\epsilon_{\alpha}$     & $L_y^*$  &  $\epsilon_{\alpha}$  &  $L_y^*$  & $\epsilon_{\alpha}$   \\
\mr
$ 60 $ & 1024  &  $2.39 \times 10^{-2}$   &  1024 & $1.31 \times 10^{-2}$   & 1152  & $6.04 \times 10^{-4}$  \\ 
  $ 75 $ &  512  &  $5.09 \times 10^{-2}$   & 1024 & $1.55 \times 10^{-2}$   & 1536 & $4.53 \times 10^{-4}$  \\ 
  $ 90 $ & 1536  &  $1.88 \times 10^{-2}$   & 1536 & $7.31 \times 10^{-3}$   & 1152  & $1.54 \times 10^{-2}$  \\   
\end{tabular}
\end{indented}
\end{table}


\subsection{Crossover length scale}

\begin{figure}[t]
  \centering
  \includegraphics[width=0.95\columnwidth]{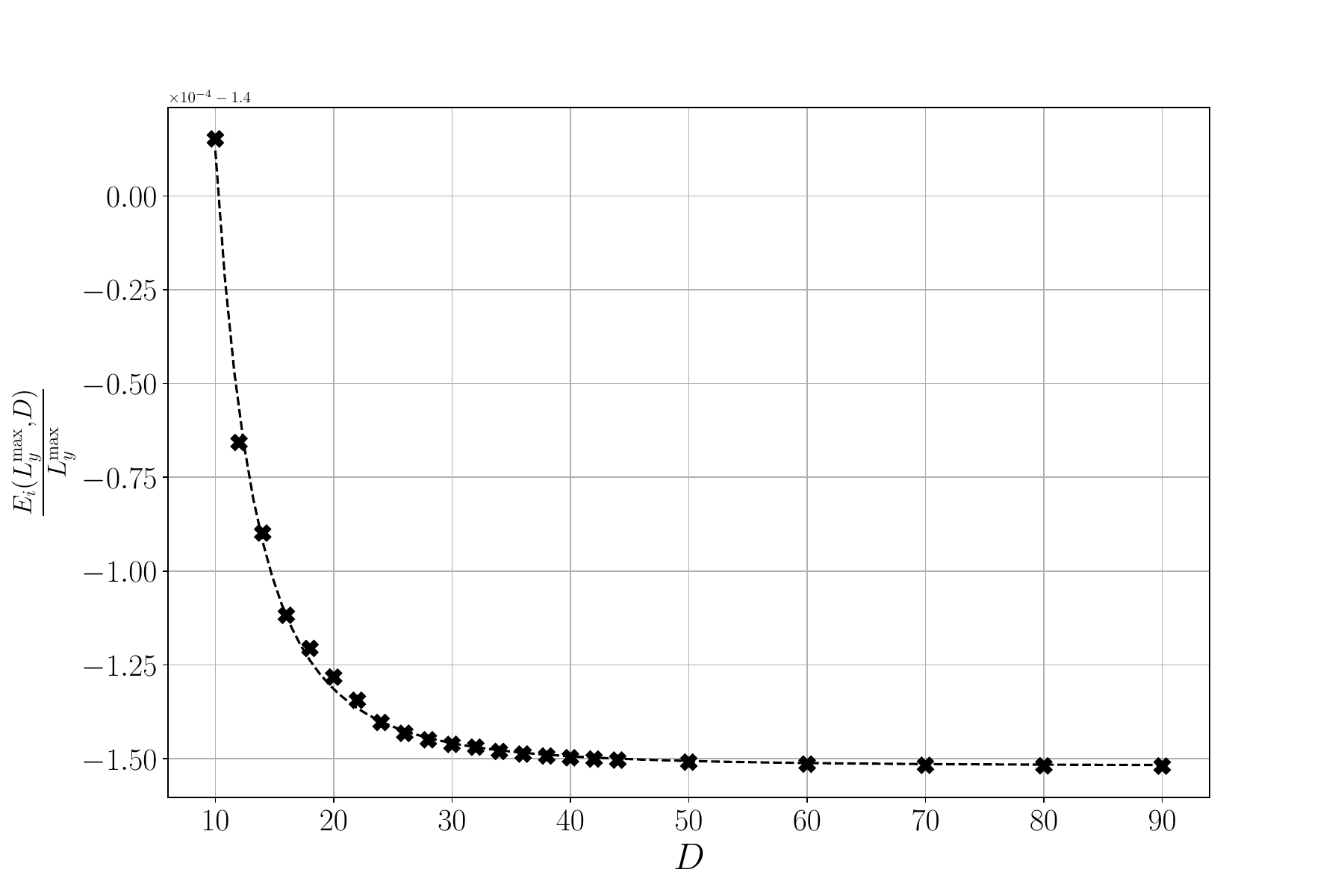}
   \caption{\label{fig:crossover}  $ \frac{E_i(T_c, L_y^{\max}, D)}{L_y^{\max}} $ as a function of $D$ for i=0 (shifted by -1.4).
   Dashed line represents fitted result based on Eq.~\ref{eq:kappa}.
   Results for $i=1,2$ look identical at this scale and are not plotted. 
   Here $L_y^{max}$ indicates the system size at which $E_i(L_y, D)/L_y$ becomes converged.} 
\end{figure}

Finally we would like to investigate if there exists a bond-dimension induced crossover length scale $\tilde{L}(D)$, which scales as $D^\kappa$
and if the exponent $\kappa$ agrees with the conventional expectation based on the conformal field theory \cite{Pirvu.2012, Ueda.2014, Stojevic:2015dj, Ueda.2017, Ueda.2020}.
From finite-size scaling theory one expects that for large enough $L_y$ one has 
\begin{equation}
  \frac{E_i(L_y)}{L_y}  = \beta_c f^\infty_c + a_i \frac{1}{L^2_y} + \cdots,
\end{equation}
where $a_i$s are non-universal. With a finite bond-dimension $D$, however, 
the length scale $L_y$ in the right hand side should be replaced by $\tilde{L}(D)$ and the equation becomes
\begin{equation}
  \label{eq:kappa}
  \lim_{L_y \rightarrow \infty} \frac{E_i(L_y, D)}{L_y}= \beta_c f^\infty_c + \tilde{a}_i \frac{1}{D^{2\kappa}} + \cdots.
\end{equation}
For a fixed $D$ we iterate the HOTRG process until $E_i(L_y, D)/L_y$ converges
and the results are plotted as a function of $D$ in Figure~\ref{fig:crossover}. 
We then fit the results to Eq.~\ref{eq:kappa}, where $\beta_c f^\infty_c$, $\tilde{a}_i$, and $2\kappa$ are fitting parameters.
We find $\beta_c f^\infty_c \approx -1.4001518(7)$ and $2\kappa \approx -3.00(4)$ consistently for $i=0,1,2$.
On the other hand, the expectation based on the conformal field theory is
\begin{equation}
  \kappa_{\mathrm{CFT}} = \frac{6}{c\left( \sqrt{ \frac{12}{c}} + 1 \right)},
\end{equation}
where $c$ is the central charge associated with the critical point.
With $c=4/5$ for the 3-state clock model, one has $2 \kappa_{\mathrm{CFT}}  \approx 3.0782$, which is close to our fitted value.
We have also tried to directly identify $\tilde{L}(D)$ similar to Ref.~\cite{Huang.2023}. 
We find that due to the finite accuracy of fitted $\beta_c f^\infty_c$,
one cannot reliably estimate $\tilde{L}(D)$ especially for larger $D$.

\section{Conclusions and Discussions\label{sec:summary}}

In summary, we use recently proposed tensor network based finite-size scaling analysis to determine the critical temperature 
and critical exponents of the two-dimensional 3-state clock model. Due to the enhanced complexity more complicated crossover behavior is observed.
However, by using the renormalization group concept and considering the relevant perturbation induced by the finite bond-dimension,
we propose a strategy to identify the perihelion of the renormalization trajectory and use that system size to extract the best estimation of critical properties for a fixed set of parameters.
Our results show that critical temperature can be estimated to extremely high accuracy.
The relative error can be reduced to the order of $10^{-7}$ with $D=70$ and $n=3$.
Consequently, the finite-size scaling analysis for other exponents can be carried out at the exact or best-estimated $T_c$.
We find that with $D=90$ and $n=3$ the relative error for the critical exponents $\nu$, $\beta$, and $\alpha$ can be reduced at least to the order of $10^{-3}$.
Moreover,  in all cases our results indicate that the errors can be systematically reduced  by increasing the bond dimension $D$ and the stacking number $n$.
It should be noted that results from $\mathbb{Z}_3$ symmetry HOTRG are reported here.
We observe that in general HOTRG without enforcing the symmetry is less stable at large $L$ and $D$, especially for quantities related to the order parameter.
When the calculations are stable, however, the results are consistent with the results reported here.
In summary the benchmark on the 3-state clock model presented in this work, and the benchmark on the Ising model reported in Ref.\cite{Huang.2023}
firmly establish that tensor network based finite-size scaling can be a powerful tool to study the phase transition and extract critical properties of 2D classical systems.
It would be interesting to investigate various models using the methods presented in this work.
For future directions, it is important and interesting to investigate how to generalize the method to study Berezinskii–Kosterlitz–Thouless transition,
weakly first order transitions, and models in higher dimensions.



\ack
We acknowledge the support by National Science and Technology Council (NSTC) of Taiwan through Grants 
No. 110-2112-M-007-037-MY3 and No. 112-2119-M-007-008.

\section*{References}
\bibliography{references}

\end{document}